\documentclass[12pt]{article}
\newtheorem{theorem}{Theorem}
\newtheorem{proposition}[theorem]{Proposition}
\newtheorem{lemma}[theorem]{Lemma}
\newtheorem{corollary}[theorem]{Corollary}
\newtheorem{definition}[theorem]{Definition}

\def\R{{\bf R}}
\def\Z{\bf Z}
\def\N{{\bf N}}

\def\C{{\bf C}}

\def\be{\begin{equation}}
\def\ee{\end{equation}}

\def\ds{\displaystyle}

\date{}
\begin{document}
\baselineskip=18pt
\title{Distributional Borel Summability for Vacuum
Polarization by an External Electric Field\footnotemark\footnotetext{Partially
supported by Universit\`a di Bologna. Funds for selected research topics.}}

\author{
Emanuela Caliceti
 \\Dipartimento di Matematica, Universit\`{a} di Bologna
\\40127 Bologna, Italy}
\maketitle
\vskip 12pt\noindent
\begin{abstract}
 { \noindent
It is proved that the divergent perturbation expansion for the
vacuum polarization by an external constant electric field in the pair
production sector is Borel summable in the distributional sense.}
\end{abstract}
\vskip 12pt\noindent
\section{Introduction and statement of the results}
\setcounter{equation}{0}%
\setcounter{theorem}{0}%
Since 1970 the standard method to deal with divergent perturbation
theory in quantum
mechanics (QM) and quantum field theory (QFT) has been Borel summability
\cite{GGS}.
For convenience of exposition, let us first recall its definition
(\cite{Bo} ; a classical  reference is \cite{Ha}; general references
especially dealing with  QM and QFT are
e.g.\cite{RS}, \cite{IZ},
\cite{ZJ}):
\begin{definition}
\label{BS}
Consider the formal power series
$\ds \sum_{n=0}^{\infty}a_nz^n$. The series
\be
\label{1.1}
B(t):=\sum_{n=0}^{\infty}\frac{a_n}{n!}t^n
\ee
is called  Borel transform of
$\sum_{n=0}^{\infty}a_nz^n$. Assume:
\begin{enumerate}
\item
 $B(t)$ has a positive radius of convergence $C>0$;
\item $B(t)$, a priori holomorphic for $|t|<C$, admits analytic
continuation at least to  a neighbourhood of the positive real axis;
\item
There is $R>0$ such that the Laplace-Borel integral
\be
\label{1.2}
f(z):=\int_0^{\infty}B(zu)\,e^{-u}\,dt
\ee
converges for $z\in C_R$ and defines an analytic function therein.
Here $C_R$ is the disk of radius $R$ tangent to the imaginary axis
at the origin, defined by $C_R:=\{z\in\C: {\rm Re}z^{-1}>R^{-1}\}$.
\end{enumerate}
Then we say
that  $\ds \sum_{n=0}^{\infty}a_nz^n$ is Borel summable to $f(z)$ for
$ z\in C_R$.
\end{definition}
{\bf Remarks}.
\begin{enumerate}
\item
If (\ref{1.1}) is inserted into (\ref{1.2}),  and summation is
formally  interchanged with integration, we see that $f(z)$ admits
the given formal power series as an asymptotic expansion as $z\to
0_+$.
An expression equivalent to (\ref{1.2}) is
\begin{eqnarray}
\label{1.2c}
\frac{1}{z}\int_0^{\infty}B(t)\,e^{-t/z}\,dt
\end{eqnarray}
\item
When the series $\ds \sum_{n=0}^{\infty}a_nz^n$ has a positive radius of
convergence the Laplace-Borel integral converges in the Borel polygon and
yields therefore the analytic continuation of the sum $f(z)$
outside the circle of convergence if this is strictly contained in
the Borel polygon; 
\item In most applications, given a
formal power series (example: a perturbation expansion) there is a
natural candidate to the sum (example, the physical solution). A
criterion is thus needed to check whether  a  formal power series 
representing the asymptotic expansion of a function is actually
Borel summable to that function. The standard one is the
Watson-Nevanlinna theorem (see
\cite{Ha}, and also
\cite{So} for a new presentation in full generality of the original
paper by Nevanlinna).
\end{enumerate}
The vacuum polarization by a constant, external (i.e., non
quantized) electromagnetic field admits a well known exact solution (\cite{Sc};
see also \cite{IZ}) which generates a divergent power series if expanded in
powers of the fine structure constant. Historically, the pure magnetic case has
been the first example where the Borel summability (of order
$2$) has been proved
\cite{Og} (see also \cite{DH}) through a direct verification of Properties 1-3
above. (Here also the Stieltjes summability holds\cite{Gr}; it entails the
convergence of the Pad\'e approximants). It is interesting to remark that the
pure electric case, which has been recently reconsidered also to discuss this
point\cite{DH}, represents instead a typical example where Borel summability
cannot hold, because the Borel transform has singularities along the positive
real axis.

To   better clarify
this point, consider the formal expansion
$\ds \sum_{n=0}^{\infty}n!z^n$. Its Borel transform is
$$
B(t)=\sum_{n=0}^{\infty}t^n=\frac{1}{1-t}\,.
$$
$B(t)$ is analytic on the whole of $\C$ except for the simple pole at
$t=1$. Consider now the function
\be
\label{1.3}
F(z):=\int_0^{\infty}\frac{e^{-t}}{1-zt}\,dt \qquad z\in\C,\;z\notin
]0,+\infty[
\ee
$F(z)$ is clearly
holomorphic in $\{z\in\C: 0<|z|;\; 0<{\rm arg}z<2\pi\}$
and its  formal expansion  at
$z=0$
is $\ds \sum_{n=0}^{\infty}n!z^n$. The non-existence of the
integral (\ref{1.3}) for
$z\in]0,+\infty[$ is due to the pole of the Borel transform $B(t)$
at $t=1$. Indeed this phenomenon  occurs whenever the coefficients
$a_n$ have a constant sign, because in that case $B(t)$ has a
singularity at $t=C$, where $C$ is the radius of convergence of
$B(t)$ (see e.g.\cite{Ti}).

If  we could perform the change of variable
(\ref{1.2c}) for $F(z)$, as we can when the conditions of
Definition 1.1 are satisfied, we could write
 \be
\label{1.4}
F(z)=\frac{1}{z}\int_0^{\infty}\frac{1}{1-t}
e^{-(t/z)}\,dt
\ee
However (\ref{1.4}) is only a formal writing because the integral
in the r.h.s. diverges for all $z\in\C$. Nevertheless (\ref{1.4})
could make sense if the Borel transform $\ds B(t)=(1-t)^{-1}$ is
regarded as an object more general than a function, for example a
distribution. More precisely, in this case we can look at the
boundary values
$\ds 
B(t\pm i0)=\frac{1}{1-t\pm i0}: t\geq 0$ of the holomorphic function $B(t)$,
as  tempered distributions:
\be
\label{1.5}
B(t\pm i0)=\frac{1}{1-t\pm
i0}
:=\lim_{\epsilon\to 0_+}\frac{1}{1-t\pm
i\epsilon}=PP\left(\frac{1}{1-t}\right)\pm i\delta(t-1)
\ee
Here
$\ds PP\left(\frac{1}{1-t}\right)$ is the Cauchy principal-value
distribution supported at $1$. Note that $B(t-i0)=\overline{B(t+i0)}$. 
Next remark that the function
\be
\label{US}
\Phi(z):=\frac{1}{z}\int_0^{\infty}B(t+i0)e^{-t/z}\,dt
\ee
exists,  is analytic for $z\in C_R$ $\forall\,R>0$, i.e. in the
half-plane ${\rm Re }z>0$, concides with $F(z)$ for $z\in\{z\in\C:
{\rm Im}z>0; {\rm Re }z>0\}$ and admits $\ds
\sum_{n=0}^{\infty}n!z^n$ as formal expansion at $z=0$. It is called
the {\it upper sum} of the series (see also the Remarks after
Definition 1.2 below). Since the divergent series is real for $z\in
[0,+\infty[$, so must be its sum provided it exists in any sense.
Therefore the natural candidate for the Borel sum is
\be
\label{dbs}
f(z):=\frac{1}{z}\int_0^{\infty}PP\left(\frac{1}{1-t}\right)e^{-t/z}\,dt=
 \frac{1}{z}\int_0^{\infty}\frac12\{B(t+i0)+\overline{B(t+i0)}\}
e^{-t/z}\,dt
\ee
In other words $\ds
f(z)=\frac12\{\Phi(z)+\overline{\Phi(\overline
z)}\}\forall\,z:{\rm Re}z>0$. In particular $f(z)={\rm Re}\Phi(z)$ for
$z\in [0,+\infty[$.

 This example shows that an extension of the
Borel  method to the case where  the Borel transform admits
singularities along the positive real axis has to allow for Borel
transforms in the sense of distributions. In turn, distributions are
particular cases of the hyperfunctions, defined as  boundary values
of holomorphic functions.  The extension, called distributional
Borel summability, has been developed in \cite{CGM}. Let us recall
here the definition and some of the main results. 
\begin{definition}
\label{DBS} 
Consider again the formal power series $\ds
\sum_{n=0}^{\infty}a_nz^n$ and its Borel
transform $B(t)$ as in Definition 1.1, with radius of
convergence $C>0$.   
Assume: 
\begin{enumerate} 
\item $B(t)$ admits
analytic continuation to the intersection of some neighbourhood of
$\R_+$ with $\C_+:=\{t\in\C: \;{\rm Im}t>0\}$; 
\item The boundary value distribution $B(t+i0)$ exists
$\forall\,t\geq 0$;
\item
Let $\ds
PP(B(t)):=\frac12\{B(t+i0)+\overline{B(t+i0)}\}$, 
$t\geq 0$. Then there exists $R>0$ such that the Laplace-Borel
integral 
\be 
\label{1.6} 
f(z):=\frac{1}{z}\int_0^{\infty}PP(B(t))
e^{-t/z}\,dt
\ee
converges for $z\in C_R$, $C_R$ as in Definition 1.1.
\end{enumerate}
Then we say that the formal power series $\ds
\sum_{n=0}^{\infty}a_nz^n$
is Borel summable in the distributional sense to
$f(z)$ for $ z\in C_R$. The distribution  $\ds PP(B(t)): t\in\R_+$ is
called distributional Borel transform of  $\ds
\sum_{n=0}^{\infty}a_nt^n$. 
\end{definition}
{\bf Remarks}.
\begin{enumerate}
\item The distribution  $\ds PP(B(t))$ coincides with the holomorphic
function $B(t)$, the Borel transform, for $0\leq t<C$;
\item
The Laplace Borel integrals
\begin{eqnarray}
\label{1.6a}
\Phi(z):=\frac{1}{z}\int_0^{\infty}B(t+ i0)
e^{-t/z}\,dt
\\
\label{1.6ab}
\overline{\Phi}(\overline{z}):=\int_0^{\infty}\overline{B(t+ i0)}
e^{-t/z}\,dt
\end{eqnarray}
exist separately in $C_R$ as analytic functions and uniquely define the
"upper" and "lower" sum, respectively.  Then
$f(z)=\{\Phi(z)+\overline\Phi(\overline z)\}/2$ for all $z\in
C_R$.In particular, $f(z)={\rm Re}\Phi(z)$, $\forall\, z\in C_R\cap\R_+$.
\item
As the ordinary Borel sum, the distributional Borel one is unique.  We note for
further reference that this method singles out also  a unique
function  with zero asymptotic
power series expansion,  the so-called 
"discontinuity",   uniquely defined by
$$
d(z):=\Phi(z)-\overline\Phi(\overline
z)=\frac{1}{z}\int_0^{\infty}\{B(t+i0)-\overline{B(t+i0)}\}
e^{-t/z}\,dt, \quad \forall\,z\in C_R
$$
In particular $d(z)=2i{\rm Im}\Phi(z)$, $\forall\,z\in C_R\cap\R_+$.
\par\noindent
In the above example we have:
$$
d(z)=\frac{1}{z}\int_0^{\infty} 2i\delta(t-1)
e^{-t/z}\,dt, \quad =\frac{2i}{z}e^{-1/z}
$$
\item
The analogue of the Watson-Nevanlinna criterion has also been
established \cite{CGM}. Its conditions have been verified to prove
the distributional Borel summability in a number of physically
interesting cases which generate constant sign divergent
perturbation expansions. Examples include the Rayleigh-Schr\"odinger
perturbation theory for Stark effect\cite{Stark} and the odd
anharmonic oscillators\cite{Odd}, which are summable to the
resonances, and  a variant of the bound state perturbation theory
for the double well quartic oscillator\cite{DW}. \end{enumerate}
Let us now proceed to state the result of this paper. Its proof is to be
described in the next section.

The effective action  for the vacuum
polarization  by a uniform electric field can be obtained as a particular case
from the Schwinger solution\cite{Sc,IZ} valid for a general external constant
electromagnetic field, and reads:
\be
\label{S1}
S(\alpha)=-\frac{1}{8\pi^2}\int_{0}^{\infty}\frac{e^{-is}}{s^3}
\left\{(2\sqrt{\pi\alpha}s)
\coth{(2\sqrt{\pi\alpha}s)}-1-\frac{4\pi\alpha s^2}{3}\right\}\,ds
\ee
Here $\alpha$  is the fine structure
constant, and without loss the electron mass $m$ and the strenght $E$ of the
field are set equal to $1$. Notice that (\ref{S1}) defines an
analytic function of $\alpha$ for $-\pi<{\rm arg}\alpha<\pi$. It can
be easily checked (see also  Lemma \ref{L1} below) that $S(\alpha)$
admits the following formal expansion in power series of $\alpha$
\begin{eqnarray} 
\label{S2}
S(\alpha)&\sim&
-\frac{1}{8\pi^2}\sum_{n=2}^{\infty}(16\pi)^nB_{2n}
\frac{(2n-3)!}{(2n)!}(-\alpha)^n=\sum_{n=2}^{\infty}a_n\alpha^n,
\\
\label{S2bis}
 a_n:&=&
-\frac{(-1)^n}{8\pi^2}(16\pi)^nB_{2n}
\frac{(2n-3)!}{(2n)!}
\end{eqnarray}
where $\{B_{2n}\}:n=0,1,\ldots$ is the sequence of the Bernoulli numbers. Now
(see e.g.\cite{GS})
\be
\label{S3}
B_0=1,
B_{2n}=2(-1)^{n+1}\frac{(2n)!}{(2\pi)^{2n}}\sum_{m=1}^{\infty}m^{-2n}, 
n=1,2,\ldots
\ee
Hence $a_n>0$ for all $n\in\N$ and $a_n\sim (2n)!$ as $n\to\infty$. Then
we can state the main result of this paper:
\begin{theorem}
\label{main}
The perturbation expansion (\ref{S2}) is Borel summable
in the distributional sense  to $\ds
\frac12\{S(\alpha)+\overline{S}(\alpha)\}={\rm Re}S(\alpha)$ for any $0\leq
\alpha<+\infty$.  More precisely $S(\alpha)$ and
$\overline{S}(\overline{\alpha})$ are the upper and lower sum of $\ds
\sum_{n=0}^{\infty}a_n\alpha^n$ for ${\rm Re}\alpha>0$,
respectively.  \end{theorem}
{\bf Remark}
\par\noindent
 The effective action $S(\alpha)$ is complex-valued, while the
perturbation expansion is real. As already remarked (and will become
evident in the course of the proof) the distributional Borel sum uniquely
determines also the imaginary part $\ds {\rm
Im}S(\alpha)=-\frac{i}{2}d(\alpha)$ , which has zero power series
expansion in
$\alpha$. This is a point of some
importance because the imaginary part is proportional to the pair
creation rate.
\section{Proof of the distributional summability}
\setcounter{equation}{0}%
\setcounter{theorem}{0}%
Consider the effective action (\ref{S1}). First of all notice that
for $0<{\rm arg}\alpha <\pi$ we can rotate the integration path in
(\ref{S1}) and choose  the half-line $\Gamma:=\{s\in\C: s=-it, 0\leq
t<+\infty\}$, i.e. the negative imaginary axis.  Now
$\coth{(ix)}=-i\cot{(x)}, \forall\,x\in A:=\{x\in\C: x\neq k\pi\,;
\forall\,k\in\Z\}$. Hence:
\begin{eqnarray}
\label{2.1}
S(\alpha)=\frac{1}{8\pi^2}\int_{0}^{\infty}\frac{e^{-t}}{t^3}
\left\{2\sqrt{\pi\alpha}t
\cot{(2\sqrt{\pi\alpha}t)}-1+\frac{4\pi\alpha
t^2}{3}\right\}\,dt
\end{eqnarray}

The proof of Theorem \ref{main} is based on the fundamental criterion for
distributional Borel summability (see \cite{CGM}, Theorem 1). Let
us report here the part relevant to our purpose.
\begin{theorem}
\label{TCGM}
Let $\ds \sum_{n-0}^{\infty}a_nz^n$ be a formal power series and $\ds
B(t)=\sum_{n-0}^{\infty}\frac{a_n}{n!}t^n$ its Borel transform. 
 Assume:
\begin{enumerate}
\item
$B(t)$ is convergent for $|t|<\rho$ for some $\rho>0$;
\item $B(t)$ 
admits an analytic continuation to the region $\Omega_\rho:=
\{t\in\C:{\rm Im }t>0; {\rm Re }t>-\rho\}$;
\item There are $A>0$, $R>0$ such that
\be
\label{2.4a}
|B(t+i\eta_0)|\leq A\eta_0^{-1}\exp{[ t/R]},
\quad\forall\,t>0,\quad
\forall \eta_0\in]0,\rho[.
\ee
\end{enumerate}
Then the boundary value distributions $\ds B(t+i0)$ and $\ds 
PP(B(t))=\frac12\{B(t+i0)+\overline {B(t+i0)}\}$ exist for all
$t\geq 0$ and the integral
\be
\label{2.5}
\frac{1}{z}\int_0^{\infty}PP(B(t))
e^{-t/z}\,dt
\ee
defines a real-analytic function $f(z)$ in $C_R$; moreover
\be
\label{2.5}
\Phi(z):=\frac{1}{z}\int_0^{\infty}B(t+i0)
e^{-t/z}\,dt
\ee
is  analytic in $C_R$ and fulfills the estimates
\be
\label{2.6}
|\Phi(z)-\sum_{n=0}^{N-1}a_nz^n|\leq
C_0c(\epsilon)^NN!|z|^N,\quad N=1,2,\ldots
\ee
uniformly in $C_{R,\epsilon}:=\{z\in C_R: {\rm arg}z\geq -\pi/2
+\epsilon\}$.
\end{theorem}
{\bf Remarks}
\begin{enumerate}
\item
$f(z)$ is the distributional Borel sum, and $\Phi(z)$ the upper
sum of $\ds \sum_{n-0}^{\infty}a_nz^n$. They are both uniquely determined
by conditions 1 and 2, together with the imaginary part $\ds
[\Phi(z)-\overline\Phi(\overline z)]/2i$.
\item The estimate (\ref{2.4a}) makes the distribution $B(t)$
locally of order $1$. However it is not a priori tempered because 
it might grow faster than any polynomial at infinity.
\end{enumerate}
Let us now proceed to apply this theorem to our case.
\begin{lemma}
\label{L1} For any $\alpha$ such that $0<{\rm arg}\alpha <\pi$ set
$\beta=\sqrt\alpha$
 and
\be
\label{Phi}
\Phi(\beta):=S(\beta^2)=\frac{1}{8\pi^2}\int_{0}^{\infty}\frac{e^{-t}}{t^3}
\left[2\sqrt{\pi}\beta t
\cot{(2\sqrt{\pi}\beta t)}-1+\frac{4\pi\beta^2
t^2}{3}\right]\,dt
\ee
Then $\Phi(\beta)$ is analytic for $0<{\rm arg}\beta<\pi/2$ and  admits
the following formal expansion in powers of $\beta$: 
\be
\label{Phiexp}
\Phi(\beta)\sim\frac{1}{8\pi^2}\sum_{n=2}^{\infty}
(-16\pi)^nB_{2n}\frac{(2n-3)!}{(2n)!}\beta^{2n}=\sum_{n=2}^{\infty}a_n\beta^{2n}
\ee
where $\{B_{2n}\}$ is the sequence of the Bernoulli numbers defined by
(\ref{S3}). 
\end{lemma}
{\bf Proof}
\par\noindent
(\ref{Phiexp}) corresponds to (\ref{S2}) with $\alpha=\beta^2$. Let us work
it out for the sake of completeness. First recall that
\be
\label{Phiexp2}
x\cot x=\sum_{n=0}^{\infty}(-1)^n\frac{2^{2n}}{(2n)!}B_{2n}x^{2n},\quad |x|<\pi
\ee
where $B_0=1$ and $B_{2n}$ is given by the expression (\ref{S3}) (see
e.g.\cite{GS}). Then
\begin{eqnarray}
\label{Phiexp3}
2\sqrt{\pi} \beta t\cot{(2\sqrt{\pi}\beta t)}=
\sum_{n=0}^{\infty}
(-1)^n\frac{2^{2n}}{(2n)!}B_{2n}2^{2n}\pi^n\beta^{2n}t^{2n}
\end{eqnarray}
Since $B_0=1$ and $B_2=1/6$ we have
\begin{eqnarray}
2\sqrt{\pi} \beta t\cot{(2\sqrt{\pi}\beta
t)}-1+\frac43\pi\beta^2t^2=
\sum_{n=2}^{\infty}(-1)^n\frac{(16\pi)^n}{(2n!)}B_{2n}\beta^{2n}t^{2n}
\end{eqnarray}
Hence
\begin{eqnarray}
\nonumber
\Phi(\beta)=\frac{1}{8\pi^2}\int_{0}^{\infty}{e^{-t}}
\sum_{n=2}^{\infty}(-1)^n\frac{(16\pi)^n}{(2n!)}B_{2n}
\beta^{2n}t^{2n-3}\,dt=
\\
\nonumber
\frac{1}{8\pi^2}\sum_{n=2}^{\infty}(-1)^n\frac{(16\pi)^n}{(2n!)}B_{2n}
\beta^{2n}\int_{0}^{\infty}{e^{-t}}t^{2n-3}\,dt=
\frac{1}{8\pi^2}\sum_{n=2}^{\infty}(-1)^n\frac{(16\pi)^n(2n-3)!}{(2n!)}B_{2n}
\beta^{2n}
\end{eqnarray}
and this concludes the proof of the Lemma.
\vskip 0.2cm
\begin{lemma}
\label{L2}
Let $D:=\{t\in\C: t\neq k\sqrt\pi/2, \,\forall\,k\in\Z\}$, and set
\be
\label{BBT}
B(t):=\frac{2\sqrt\pi t\cot{(2\sqrt\pi t)}-1+4\pi t^2/3}{8\pi^2t^3},
\quad \forall\,t\in D
\ee
Then $B(t)$ is clearly analytic in $D$ with simple poles at
$t=k\sqrt\pi/2$, $k\in\Z$. Moreover:
\be
\label{BBTT}
B(t)=\sum_{n=2}^{\infty}\frac{a_n}{(2n-3)!}t^{2n-3}, \quad \forall\,t:
|t|<\frac{\sqrt\pi}{2}
\ee
i.e. $B(t)$ is the Borel transform of $\ds
\sum_{n=2}^{\infty}{a_n}\beta^{2n-3}$.
\end{lemma}
\noindent
{\bf Proof}. 
To obtain (\ref{BBTT}) we proceed as in the previous Lemma using
(\ref{Phiexp2}).  More precisely:
\begin{eqnarray*}
B(t)=\frac{1}{8\pi^2}\sum_{n=2}^{\infty}(-1)^n\frac{2^{2n}}{(2n!)}B_{2n}
2^{2n}\pi^{n}t^{2n-3}=\sum_{n=2}^{\infty}\frac{a_n}{(2n-3)!}t^{2n-3},
\;|t|<\sqrt\pi/2 .
\end{eqnarray*}
\begin{proposition}
\label{stima}
Let $\ds \sum_{n=2}^{\infty}{a_n}\beta^{2n-3}$ be the formal power
series whose coefficients $a_n$ are defined by (\ref{S2}, \ref{S2bis})
(see also (\ref{Phiexp})), and let $B(t)$ be its Borel transform
(\ref{BBT}) with the expansion (\ref{BBTT}). Then $B(t)$ satisfies
the hypotheses of Theorem \ref{TCGM}.
\end{proposition}
{\bf Proof} 
By Lemma \ref{L2} Conditions 1 and 2 of Theorem \ref{TCGM} are satisfied
with $\rho=\sqrt\pi/2$. As far as Condition 3 is concerned, we will prove
it in the following stronger version: for any $R>0$ there is $A>0$ such
that
\be
\label{2.4}
|B(t+i\eta_0)|\leq A\eta_0^{-1}\exp{(t/R)}, \quad\forall\,t>0,\quad
\forall \eta_0\in]0,\sqrt\pi/2[.
\ee
To this end first let  $\ds 0<\delta<\sqrt\pi/2$ be fixed. Then the
function $B(t+i\eta_0)$ is continuous on the compact set
$K:=\{z=t+i\eta_0\in\C: |t|\leq\delta; 0\leq \eta_0\leq
\sqrt\pi/2\}$. Hence $B(t+i\eta_0)$ is bounded in $K$ by some constant
$c>0$ and we can write
$$
|B(t+i\eta_0)|\leq c\leq c\frac{\sqrt\pi}{2}\eta_0^{-1}\exp{( t/R)}
$$
where the second inequality holds because $\ds
\frac{\sqrt\pi}{2}\eta_0^{-1}\geq 1$ and obviously we can choose $R$ as
large as we like . Hence it suffices to prove (\ref{2.4}) for $t>\delta$
and
$\eta_0\in]0,\sqrt\pi/2[$. Now for
$t>\delta$ the term $\ds \frac{1}{6\pi|t+i\eta_0|}$, which comes from the
third summand in (\ref{BBT}) where we have replace $t$ by $t+i\eta_0$, 
can be estimated as follows:
$$
\frac{1}{6\pi|t+i\eta_0|}= \frac{1}{6\pi\sqrt{t^2+\eta_0^2}}\leq
\frac{1}{6\pi\delta}
$$
for $t>\delta$. Thus, this term trivially fulfills (\ref{2.4}) with $R$
as large as we like. Therefore we can restrict our attention to the term
\be
\label{2.5}
B_1(t+i\eta_0):=2\sqrt{\pi}(t+i\eta_0)\cot{\{2\sqrt{\pi}
(t+i\eta_0)\}}-1 
\ee
because the denominator $\ds \frac{1}{8\pi^2|t+i\eta_0|^3}$ is bounded by
$\ds \frac{1}{8\pi^2}
\delta^{-3}$ for $t>\delta$.  Consider now the well known
expansion (see e.g.\cite{GS})
\be
\label{2.6}
x\cot{x}=1+2x^2\sum_{n=1}^{\infty}\frac{1}{x^2
-n^2\pi^2}
\ee
Then, for any $R>0$, we have to find $A>0$ such that
\be
\label{2.7}
|B_1(t+i\eta_0)|=4\pi|t+i\eta_0|^2\left|\sum_{n=1}^{\infty}
\frac{1}{4\pi(t+i\eta_0)^2
-n^2\pi^2}\right|\leq A \eta_0^{-1}\exp{ (t/R)}
\ee
$\forall\, t>\delta$, $\eta_0\in ]0,\sqrt\pi/2[$. First remark that
$$
4\pi|t+i\eta_0|^2=4\pi(t^2+\eta_0^2)\leq 4\pi (t^2+1)\leq
 C_1 \exp{(t/R)}
$$
for a suitable constant $C_1>$ and $R>0$ arbitrarily large. Moreover 
one has
\begin{equation}
\label{2.8}
\left|\sum_{n=1}^{\infty}\frac{1}{4(t+i\eta_0)^2
/\pi-n^2}\right|\leq
\sum_{n=1}^{\infty}\frac{1}
{\sqrt{\{n^2-4(t^2-\eta_0^2)/\pi\}^2+64\eta_0^2t^2/\pi^2}} 
\end{equation}
If $t\leq \eta_0$ the right hand side of eq.(\ref{2.8}) can be bounded by $\ds
\sum_{n=0}^{\infty}\frac{1}{n^2}=\pi^2/6$. Thus (\ref{2.7}) holds for
$R>0$ arbitrarily large by suitably choosing $A>0$. We are thus left with
the case $t>\eta_0$, $t>\delta$, $\eta_0\in ]0,\sqrt\pi/2[$. In this
case, setting $\ds Q(t,\eta_0):=[2\sqrt{t^2-\eta_0^2}]/\sqrt\pi]+1$
($[x]$=greatest integer $\leq x$, $x\in\R$) we can estimate the
r.h.s. of (\ref{2.8}) as follows:
\begin{eqnarray}
\nonumber
\sum_{n=1}^{\infty}\frac{1}
{\sqrt{\{n^2-4(t^2-\eta_0^2)/\pi\}^2+64\eta_0^2t^2/\pi^2}}
\\
\nonumber
\leq \sum_{n=1}^{Q(t,\eta_0)}\frac{1}
{\sqrt{\{n^2-4(t^2-\eta_0^2)/\pi\}^2+64\eta_0^2t^2/\pi^2}}
\\
\nonumber
+\sum_{n=Q(t,\eta_0)+1}^{\infty}\frac{1}
{\sqrt{\{n^2-4(t^2-\eta_0^2)/\pi\}^2+64\eta_0^2t^2/\pi^2}}
\\
\label{2.9}
\leq
\sum_{n=1}^{Q(t,\eta_0)}\frac{\pi}{8t\eta_0}+
\sum_{n=Q(t,\eta_0)+1}^{\infty}\frac{1}{n^2-4(t^2-\eta_0^2)/\pi}
\end{eqnarray}
The last inequality is a consequence of the positivity of $\ds 
n^2-4(t^2-\eta_0^2)/\pi$ for $\ds n\geq Q(t,\eta_0)+1$. Now the first
summand in (\ref{2.9}) can be bounded by:
\begin{eqnarray}
\label{2.10}
\frac{\pi}{8t\eta_0}(\frac{2}{\sqrt\pi}\sqrt{t^2-\eta_0^2}+1)\leq 
\frac{\sqrt\pi}{4}\eta_0^{-1}+\frac{\pi}{8t\eta_0}
\end{eqnarray}
and clearly satisfies (\ref{2.4}) recalling that $t>\delta$.
Concering the second term in (\ref{2.9}) we have
\begin{eqnarray}
\nonumber
\sum_{Q(t,\eta_0)+1}^{\infty}\frac{1}{n^2-4(t^2-\eta_0^2)/\pi}
\leq
\int_{Q(t,\eta_0)+1}^{\infty}\;\frac{dx}{x^2-4(t^2-\eta_0^2)/\pi}
\\
\label{2.11}
+\frac{1}{(Q(t,\eta_0)+1)^2-4(t^2-\eta_0^2)/\pi}
\end{eqnarray}
where the inequality
follows by the well known comparison theorem between series with positive
terms and generalized integrals.  Since $[x]\leq x\,\forall\,x\geq
0$, recalling the definition of $Q(t,\eta_0)$, we can write
\be
\label{2.12}
\sum_{Q(t,\eta_0)+1}^{\infty}\frac{1}{n^2-4(t^2-\eta_0^2)/\pi}\leq
\int_{\frac{2}{\sqrt\pi}\sqrt{t^2-\eta_0^2}+2}^{\infty}\;\frac{dx}{x^2-4(t^2-\eta_0^2)/\pi}+\frac13
\ee
because $\ds [x]^2=[x^2]$ and $\ds [x]+1-x>0$, $\forall\,x>0$. Since
the additive factor $1/3$ can be trivially absorbed in the constants, 
it is
enough to estimate the integral in (\ref{2.12}). One has:
\begin{eqnarray*}
\int_{\frac{2}{\sqrt\pi}\sqrt{t^2-\eta_0^2}+2}^{\infty}\;\frac{dx}
{x^2-4(t^2-\eta_0^2)/\pi}=
\frac14\sqrt{\frac{\pi}{t^2-\eta_0^2}}\ln{\left(1+
\frac{2}{\sqrt{\pi}}\sqrt{t^2-\eta_0^2}\right)}.
\end{eqnarray*}
Given  $R>0$  arbitrarily large the existence of a constant $A>0$ such
that
$$
\frac14\sqrt{\frac{\pi}{t^2-\eta_0^2}}\ln{\left(1+
\frac{2}{\sqrt{\pi}}\sqrt{t^2-\eta_0^2}\right)}\leq
A\eta_0^{-1}e^{t/R} 
$$
$\forall\,t>\delta$, $t>\eta_0$, $\forall\,\eta_0\in]0,\sqrt\pi/2[$ is now
obvious. This concludes the proof of the Proposition.
\begin{corollary}
\label{cor}
In the notations of Lemmas \ref{L1},\ref{L2} and Proposition \ref{stima}
the boundary value distributions $B(t+i0)$ and $PP(B(t))$ exist for
all $t\geq 0$. Moreover the integral
\be
\label{2.13}
\frac1{\beta}\int_0^{\infty}B(t+i0)e^{-t/\beta}\,dt
\ee
defines an analytic function on $D_1:=\{\beta\in\C: {\rm Re}\beta >0\}$,
coinciding with $F(\beta):=\beta^{-3}\Phi(\beta)$ $\forall\,\beta: 0<{\rm
arg}\beta <\pi/2$.
\noindent
Equivalently: $\ds \sum_{n=0}^{\infty}a_n\beta^{2n-3}$ is Borel summable
in the distributional sense; its upper sum is $\beta^{-3}\Phi(\beta)$
while the distributional sum is
\be
\label{2.14a}
f(\beta):=\frac1{\beta}\int_0^{\infty}PP(B(t))e^{-t/\beta}\,dt
\ee
$\forall\,\beta\in D_1$; moreover
\be
\label{2.14b}
f(\beta)=\frac1{2}\{\beta^{-3}\Phi(\beta)+(\overline\beta)^{-3}
\overline\Phi(\overline\beta)\}, \quad \beta\notin \R_+.
\ee
\end{corollary}
{\bf Proof}. The first assertion follows from Theorem \ref{TCGM} in view
of Proposition \ref{stima}, which also guarantee the existence of
(\ref{2.13}) in $C_R=\{\beta\in\C: {\rm Re }\beta^{-1}>R^{-1}\}$
$\forall\,R>0$. By the same results it suffices now to show that
(\ref{2.13}) coincides with $\ds \beta^{-3}\Phi(\beta)$ for $0<{\rm
arg}\beta<\pi/2$. Indeed we have
\begin{eqnarray}
\nonumber
\frac1{\beta}\int_0^{\infty}B(t+i0)e^{-t/\beta}\,dt=
\frac1{\beta}\lim_{\epsilon\to
0_+}\int_0^{\infty}B(t+i\epsilon)e^{-t/\beta}\,dt=
\\
\nonumber
\frac1{8\pi^2\beta}\lim_{\epsilon\to
0_+}\int_0^{\infty}e^{-t/\beta}(t+i\epsilon)^{-3}\{2\sqrt\pi
(t+i\epsilon)\cot{(2\sqrt\pi(t+i\epsilon))}-1+\frac43\pi(t+i\epsilon)^2\}\,dt
\end{eqnarray}
Now, performing the change of variables $t=s\beta$, $s\in\Gamma_1:=\{s:
s=\frac{t}{\beta}, 0\leq t<\infty\}$ we obtain:
\begin{eqnarray}
\nonumber
\frac1{\beta}\int_0^{\infty}B(t+i0)e^{-t/\beta}\,dt=\qquad\qquad\qquad\qquad
\\
\nonumber
\frac{1}{8\pi^2\beta^3}
\int_{\Gamma_1}s^{-3}e^{-s}\{(2\sqrt\pi \beta
s)\cot{(2\sqrt\pi \beta s)}-1+\frac43\pi\beta^2s^2\}\,ds=
\\
\label{2.18}
\frac{1}{8\pi^2\beta^3}
\int_{0}^{\infty}t^{-3}e^{-t}\{(2\sqrt\pi \beta
t)\cot{(2\sqrt\pi \beta t)}-1+\frac43\pi\beta^2t^2\}\,dt
\end{eqnarray}
where the last equality follows from the analyticity of the integrand
(in the variable $s$) in a sector containing $\R_+$, if
$\beta\notin\R_+$. Now by  (\ref{Phi}) the  integral (\ref{2.18}) is
precisely $\ds
\beta^{-3}\Phi(\beta)$, and this concludes the proof of the
Corollary. \par\noindent
{\bf Remark}.  Notice that in the representations 
(\ref{2.14a},\ref{2.14b}) we had to exclude $\beta\in\R$ because by
(\ref{Phi}) $\Phi(\beta)$ is not defined for $\beta\in\R$.
\vskip 0.2cm\noindent
Now, multiplying by $\beta^3$ the functions
$F(\beta)=\beta^{-3}\Phi(\beta)$ and $f(\beta)$ as well as the  formal
series $\ds \sum_{n=2}^{\infty}a_n\beta^{2n-3}$ we immediately conclude
\begin{corollary}
\label{cor1}
The formal power series $\ds \sum_{n=2}^{\infty}a_n\beta^{2n}$ is Borel
summable in the distributional sense $\forall\,\beta: {\rm Re}\beta >0$.
Its upper sum is $\Phi(\beta)$, $\forall\,\beta\notin\R_+$, and its
distributional sum is $\beta^3f(\beta)$. For $\beta\notin\R_+$ one has
$\ds \beta^3f(\beta)=\frac12\{\Phi(\beta)+\overline\Phi(\overline\beta)\}$. 
\end{corollary} 
\vskip 0.2cm\noindent
{\bf Proof of Theorem \ref{main}} 
\vskip 0.2cm\noindent
>From Corollary (\ref{cor1}) we obtain
\be
\label{2.19}
\Phi(\beta)=\beta^2\int_0^{\infty}B(t+i0)e^{-t/\beta}\,dt, \quad
\forall\,\beta: 0<{\rm arg}\beta<\frac{\pi}{2}
\ee
where the r.h.s. is the upper sum of $\ds
\sum_{n=2}^{\infty}a_n\beta^{2n}$, $\ds \forall\,\beta:-\frac{\pi}{2}<
{\rm arg}\beta<\frac{\pi}{2}$. Now, setting $\beta=\sqrt\alpha$ and using
(\ref{2.1}) to represent $S(\alpha)$ we have
\begin{eqnarray}
\label{2.20}
S(\alpha)=\sqrt\alpha\int_0^{\infty}B(t+i0)e^{-t/\sqrt\alpha}\,dt
\end{eqnarray}
and
$$
\overline
S(\overline\alpha)=\sqrt{\alpha}\int_0^{\infty}\overline{B(t+i0)}
e^{-t/\sqrt{\alpha}}\,dt
$$
$\forall\,\alpha: 0<{\rm arg}\alpha <\pi$ (notice  that (\ref{2.1})
actually  defines $S$ as a holomorphic function of $\alpha$ for $0<{\rm
arg}\alpha <\pi$). On the other hand the original representation
(\ref{S1}) for
$S(\alpha)$, namely
$$
S(\alpha)=-\frac{1}{8\pi^2}\int_{0}^{\infty}\frac{e^{-is}}{s^3}
\left\{(2\sqrt{\pi\alpha}s)
\coth{(2\sqrt{\pi\alpha}s)}-1-\frac{4\pi\alpha s^2}{3}\right\}\,ds
$$
defines $S$ as a holomorphic function of $\alpha$ for $\ds
-{\pi}<{\rm arg}\alpha <{\pi}$. Thus, it represents 
an analytic continuation of the l.h.s. of (\ref{2.20}) across the
positive real axis because
$S(\alpha)$ as represented by (\ref{2.20}) and (\ref{2.1}) coincide for 
$\ds 0<{\rm arg}\alpha <{\pi}$. Since the r.h.s. of (\ref{2.20})
is also an analytic function of $\alpha$ for $\ds
-{\pi}<{\rm arg}\alpha <{\pi}$ we can write
\begin{eqnarray}
\nonumber
S(\alpha)=-\frac{1}{8\pi^2}\int_{0}^{\infty}\frac{e^{-is}}{s^3}
\left\{(2\sqrt{\pi\alpha}s)
\coth{(2\sqrt{\pi\alpha}s)}-1-\frac{4\pi\alpha s^2}{3}\right\}\,ds
\\
\label{2.22}
=\sqrt\alpha\int_{0}^{\infty}B(t+i0)e^{-t/\sqrt\alpha}\,dt,
\quad
\forall\,\alpha: -{\pi}<{\rm arg}\alpha <{\pi}
\end{eqnarray}
Thus $S(\alpha)$ is the upper sum of $\ds \sum_{n=2}^{\infty}a_n\alpha^n$
and the distributional Borel sum is given by
\be
\label{2.23}
\frac12\{S(\alpha)+\overline
S(\overline\alpha)\}=\sqrt\alpha
\int_{0}^{\infty}PP(B(t))e^{-t/\sqrt\alpha}\,dt, \quad
\forall\,\alpha: -{\pi}<{\rm arg}\alpha <{\pi}
\ee
In particular, the distributional Borel sum is ${\rm Re}S(\alpha)$ for
$\alpha >0$. This concludes the proof of Theorem \ref{main}.
\vskip 12pt\noindent
{\bf Remarks}. 
\begin{enumerate}
\item It follows by the above theorem that the distributional Borel
summability uniquely determines also
$$
{\rm Im}S(\alpha)=\frac{1}{2i}[S(\alpha)-\overline
S(\alpha)], \quad \alpha\in\R_+.
$$
Moreover ${\rm Im}S(\alpha)$ is proportional to the pair-production rate. Its
explicite expression is \cite{IZ}:
$$
{\rm
Im}S(\alpha)=\frac{1}{8\pi^3}\sum_{n=1}^{\infty}\frac1{n^2}
\exp{\left(-\frac{n\pi}{\alpha}\right)}
$$
and has zero asymptotic expansion in $\alpha$. 
\item
Strictly speaking, the representations (\ref{2.22},\ref{2.23})
yield the distributional Borel-Leroy sum of order $2$ (see
\cite{CGM}, Theorem 3) of the divergent perturbation expansion
(\ref{S2}). That definition is completely equivalent (\cite{CGM}) to 
ordinary summability in the variable $\beta=\sqrt\alpha$. We have
preferred to proceed in this last way for convenience of exposition.
\end{enumerate}

\end{document}